# Role of Orientational Disorder in ZSM-22 in the Adsorption of SO$_2$

Sadique Vellamarthodika [1] and Siddharth Gautam [2,*]

[1]Positron Foundation for Science and Innovation, CITTIC, Cochin University of Science and Technology, Cochin, Kerala 682022, India

[2]School of Earth Sciences, The Ohio State University, 125 S Oval Mall, Columbus 43210 USA

**Abstract**

Computational studies addressing the adsorption of fluids in nano-porous materials mostly use ideal single crystal models of the adsorbent. While a few recent studies have tried to address the effects of inter-crystalline spacing on the adsorption of fluids in polycrystalline models of nano-porous materials, the effects of the orientational disorder in the polycrystalline adsorbent remain unexplored. Here we report the adsorption of SO$_2$ – an industrially and environmentally important gas – in ZSM-22, a zeolite characterized by straight channel like pores. The simple pore geometry of ZSM-22 helps us make polycrystalline models of the adsorbent with different degrees of orientational disorder. Using grand canonical Monte Carlo (GCMC) simulations, we obtain the adsorption isotherms of SO$_2$ in ZSM-22 with different inter-crystalline spacings and degree of orientational disorder. Introducing inter-crystalline space is found to enhance the adsorption capacity, with a larger inter-crystalline space leading to higher adsorption. Increasing the orientational disorder of the adsorbent is found to enhance the adsorption capacity too. However, the effects of orientational disorder become weaker when the inter-crystalline space is widened. This weakening of the effect of orientational disorder is a result of an interplay between the width of the inter-crystalline space and the strength of guest-guest interactions.

## 1. Introduction

Molecular simulations are a relatively cheap and safe way to understand material behavior at the molecular scale [1]. Molecular dynamics (MD) and Monte Carlo (MC) simulations are often employed to study the behavior of confined fluids [2]. For simplicity and ease of computation, these simulations often use ideal single crystal models of the nano-porous confining medium. In reality however, most experiments are carried out on polycrystalline samples. Although amorphous silica pore models have been closer to reality with disordered models addressing non-crystallinity [3-4] and even irregular pore shapes [5-6], zeolite materials are often modeled as perfect crystals. This difference in the models used in the simulations and the more disordered samples used in the experiments has not been addressed in much details. Earlier attempts at introducing imperfection in zeolitic models involved systematically varying the cation content in cation-exchanged models of zeolites with finite Si/Al ratios [7, 8]. Recently, attempts have been made at incorporating polycrystallinity and disorder in the pore network of ZSM-5 to address the effects of inter-crystalline spacing [9] and pore connectivity [10, 11] on the behavior of confined fluids. Both pore connectivity and presence of inter-crystalline spacing have been found to affect the behavior of confined fluid significantly. While polycrystallinity in these studies was modeled by means of inserting an inter-crystalline space between crystallites with perfect ordering, the individual crystallites themselves were arranged in a perfect order with all crystallites oriented in the same direction. For systems with pores in the form of channels, the relative orientation of the crystallites can affect the behavior of the confined fluid and therefore it is imperative to systematically study the effects of orientational disorder in the polycrystalline models of nano-porous materials due to crystallites disoriented with respect to each other.

ZSM-22 is an all-silica zeolite characterized by straight parallel channel like pores of sub-nanometer size [12]. With a simple geometry of the individual pore channels and their network, ZSM-22 provides a good model system to systematically study the effects of orientational disorder in the polycrystalline model on the behavior of confined fluid.

$SO_2$ is an industrially and environmentally important gas [13]. Several industrial as well as natural processes release $SO_2$ in the atmosphere where it might undergo reactions to produce toxic species [14]. Given the harmful effects this atmospheric $SO_2$ might have on life in general, it is important to capture and store this gas at the source. Adsorption in nano-porous materials is a good strategy for capture and storage of gases [15]. In particular zeolites provide a promising adsorbent for capturing and storing different gases [16]. ZSM-22, with its sub-nanometer channel like pores can be a good candidate for capturing $SO_2$.

Here we study the adsorption of $SO_2$ in polycrystalline models of ZSM-22 using grand canonical Monte Carlo (GCMC) simulations. Polycrystalline models of ZSM-22 are made using perfect crystallites of ZSM-22 separated from each other by inserting inter-crystalline spacing of different widths. The orientation of different crystallites is changed with respect to each other to obtain polycrystalline models with different degrees of orientational disorder. Both inter-crystalline spacing and orientational disorder in the adsorbent is found to affect the sorption of $SO_2$. In section 2, we detail the molecular models and simulation protocols used in the study. This is followed by section 3 which lists all the results whose implications are discussed in section 4. We conclude by enlisting important inferences drawn from the study in section 5.

## 2. Materials and Methods

### 2.1. Models of polycrystalline ZSM-22

Atomic coordinates provided by Kokotailo et al [12] were used to make models of the ZSM-22 adsorbent. To start with, a unit cell was replicated 2x1x3 times to make a single crystallite using the visualization software VESTA [17]. Eight such crystallites were then arranged in a 2x2x2 lattice separated from each other by an inter-crystalline space of 4 or 8 Å to stand for small and large inter-crystalline space respectively. A vacuum equal to half the inter-crystalline gap was left on all six sides of the supercell made in this way. With periodic boundary conditions applied, this supercell thus constitutes an infinitely extended system consisting of crystallites separated from each other by a fixed and uniform inter-crystalline space in all three directions. This resulted in two polycrystalline models with different inter-crystalline spacings labeled respectively 'small gap (SG)' and 'large gap (LG)'. In addition, a model was made with no inter-crystalline gap between the crystallites. This is a perfectly ordered single crystal model of ZSM-22 and is labeled 'no gap (NG)'. To add orientational disorder to the SG and LG models, one out of eight crystallites were rotated in their position by 90 degrees at a time. A series of models with different degrees of orientational disorder (OD) were thus obtained by rotating between one to four out of eight crystallites in the LG or SG systems. We note that although the individual crystallites were not cubic, their dimensions were such that this rotation did not significantly alter the inter-crystalline spacing. The resulting models are named SG-OD$n$ and LG-OD$n$ respectively with $n$ denoting the number of crystallites that are rotated about their position. In all 11 different models of ZSM-22 were made for the simulations. These are schematically shown in Figure 1 and are listed in Table 1.

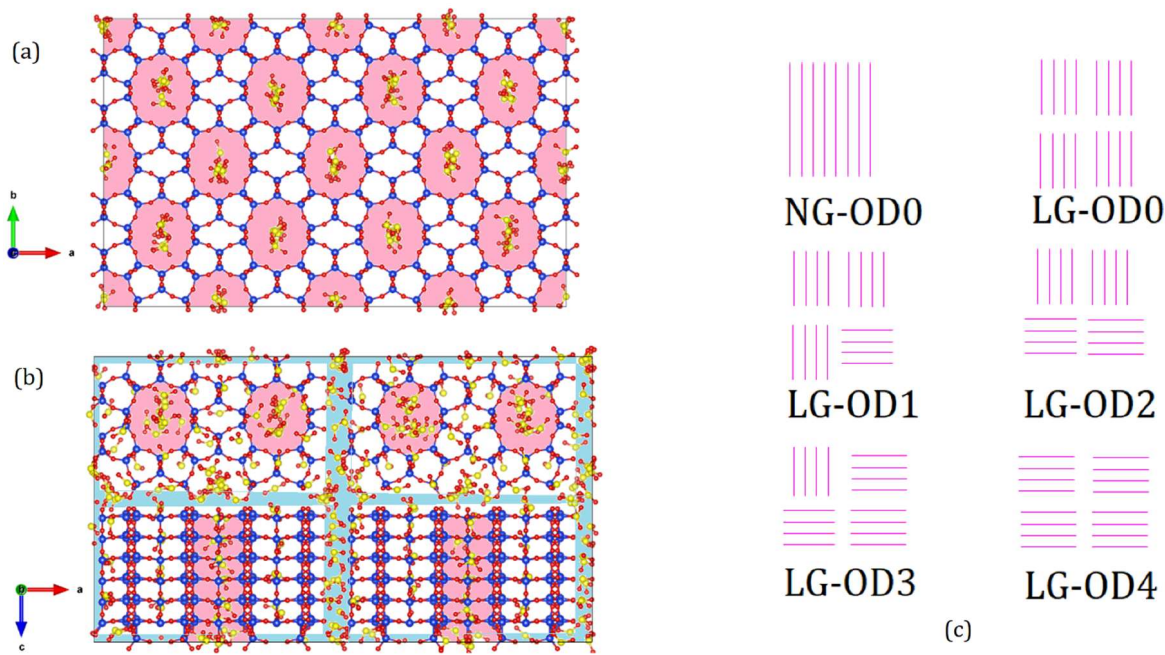

**Figure 1.** (a) Simulation snapshot for the perfect crystal system ZSM-22 (NG-OD0) with $SO_2$ adsorbed at a partial gas pressure of 100 atm. Si, S, and O atoms are shown as spheres of blue, red and yellow color respectively. (b) Simulation snapshot of $SO_2$ adsorbed in the disordered ZSM-22 system SG-OD4 at 100 atm. Individual crystallites are separated from each other by inter-crystalline spacing shown in light blue whereas channel-like pores in the crystallites are marked in light magenta. (c) Schematic representation of the pore network in some ZSM-22 models. Out of a total of 8 crystallites, only 4 that are visible from front are shown. Channel-like pores are shown as single magenta line. In NG-OD0, all crystallites are stacked on top of each other so that channels belonging to two vertically adjacent crystallites form one single channel. In LG-OD0 (and similarly, in SG-OD0 not shown here) individual crystallites have space between them resulting in a discontinuation of channels. In LG-OD1, one crystallite can be seen disoriented (horizontal instead of vertical lines in the bottom right crystallite). The number of crystallites disoriented in this fashion increases in LG-OD2, LG-OD3 and LG-OD4 resulting in an increase in orientational disorder. Note that the crystallites at the back (not shown) retain their original orientation. Thus, in LG-OD4, all 4 crystallites visible in front have an orientation different from the four crystallites at the back (not visible).

*2.2. Force-Fields*

Sokolić et al [18] proposed four effective sets of pair potential to model $SO_2$. Later, Ribeiro [19] used one of these sets of parameters to obtain thermodynamic, structural and dynamical properties of liquid $SO_2$ obtaining good agreement with experimental data. In the present study we use the parameters employed by Ribeiro [19]. Sun and Han [20] have used these parameters for $SO_2$ to simulate $SO_2$ in MFI and 4A zeolites. $SO_2$ molecule in this formalism is represented by a sulfur (S) atom rigidly bonded to two oxygen ($O_s$) atoms with a fixed bond length of 1.4321 Å each. $O_s$-S-$O_s$ angle is fixed at 119.5$^0$. To model, the interactions of $SO_2$ with ZSM-22 we used the CLAYFF [21] force-field parameters to represent silicon and oxygen atoms. All intermolecular interactions thus had a van der Waals component represented by Lennard-Jones type of interactions and an electrostatic component represented by Coulombic interactions. All the corresponding potential parameters used in the study are listed in Table 2. Cross-term parameters were calculated using the Lorentz-Berthelot mixing rules. The long-range electrostatic interactions were calculated using the Ewald sum method [2]. We used an interaction cut-off of 14 Å.

**Table 1.** Models of ZSM-22 used in the simulations.

| Model | Inter-crystalline Spacing (Å) | Number of disoriented crystallites |
|---|---|---|
| NG-OD0 | 0 | 0 |
| SG-OD0 | 4 | 0 |
| SG-OD1 | 4 | 1 |
| SG-OD2 | 4 | 2 |
| SG-OD3 | 4 | 3 |
| SG-OD4 | 4 | 4 |
| LG-OD0 | 8 | 0 |
| LG-OD1 | 8 | 1 |
| LG-OD2 | 8 | 2 |
| LG-OD3 | 8 | 3 |
| LG-OD4 | 8 | 4 |

**Table 2.** Force-field parameters of atoms belonging to ZSM-22 (first two rows) and $SO_2$ (bottom two rows) used in the simulations. The parameters for ZSM-22 and $SO_2$ are taken from ref. 20 and 19 respectively.

| Atom[1] | Charge (q/e) | ε (kJ) | σ (Å) |
|---|---|---|---|
| Si | +2.100 | 0.000008 | 3.301 |
| O | -1.050 | 0.650198 | 3.166 |
| S | +0.470 | 1.283740 | 3.585 |
| $O_s$ | -0.235 | 0.517990 | 2.993 |

[1] To distinguish oxygen atoms belonging to ZSM-22 and $SO_2$ we label the former as O and later as $O_s$.

## 2.2. Simulations

Grand canonical Monte Carlo (GCMC) simulations were carried out using DL-Monte [22]. To start with one $SO_2$ molecule was kept at the center of the simulation cell. GCMC simulations were then carried out at a given partial pressure of $SO_2$. During the simulation, the guest molecules, i.e., $SO_2$, could be inserted/deleted, translated, or rotated with respective probabilities of 0.5, 0.25, and 0.25, while all ZSM-22 atoms were kept rigid. 2 million Mote Carlo simulation steps were used to obtain the number of $SO_2$ adsorbed in the model ZSM-22 consistent with a given gas partial pressure and temperature. Out of this the first 500000 steps were discarded for equilibration. All simulations were carried out at 300 K. In all 242 simulations were carried out.

## 3. Results

### 3.1. Adsorption Isotherms

In Figure 2, we show the adsorption isotherms of $SO_2$ in the 11 models of ZSM-22 listed in Table 1. The effects of inter-crystalline spacing in models with perfect orientational order is shown in Figure 2a. Introducing small crystalline gaps to NG-OD0 as in SG-OD0 enhances the adsorption capacity of ZSM-22 for $SO_2$. Moreover, as this inter-crystalline gap widens in LG-OD0, the adsorption capacity increases further. In Figure 2b, the effects of the orientational disorder can be seen in SG-OD$n$ models. With an increase in orientational disorder, the amount of $SO_2$ adsorbed in ZSM-22 increases progressively at all gas partial pressures. A similar increase in the adsorption amounts with an increase in orientational disorder is seen in models with larger gaps (LG-OD$n$) at intermediate pressures (Figure 2c). However, this increase is much smaller as compared to that in SG-OD$n$. Moreover, in LG-OD$n$ models, the relation between orientational disorder and the adsorption amounts reverses at higher partial pressures of $SO_2$, whereas at low pressures the amount of adsorption is independent of orientational disorder.

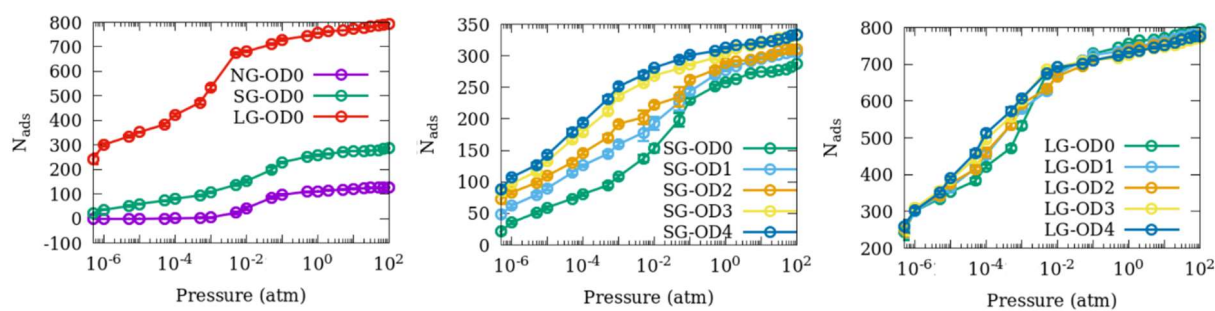

**Figure 2**. Adsorption isotherms of $SO_2$ at 300 K in ZSM-22 models with (a) different inter-crystalline spacings and uniform orientation, (b) small inter-crystalline space and different degrees of orientational disorder and (c) large inter-crystalline space and different degrees of orientational disorder.

### 3.2. Pair distribution functions

In Figure 3 we show the pair distribution functions of the pair of atoms belonging to ZSM-22 and $SO_2$. For clarity we present the data only for the models NG-OD0, SG-OD0, LG-OD0, SG-OD4 and LG-OD4. These are representatives of the three gaps and extreme cases of orientational disorder investigated. PDFs for NG-OD0 model are shown in Figures 3a and 3b. Both the S-Si and S-O pair representing the guest-zeolite pairs exhibit several relatively sharp peaks indicating a significant order. In contrast, the S-S pair representing the guest-guest interaction shows a strong sharp peak followed by a rather weak peak and a pair of two closely placed peaks. The first S-S peak is located at distances shorter than the position of the first S-O and S-Si peaks suggesting a stronger guest-guest interaction compared to the guest-zeolite interactions. S atom from the guest interacts closely with O from the zeolite compared to Si. Similar behavior is shown by the PDF corresponding to the oxygen ($O_s$) atom from the guest (Figure 3b). In case of $O_s$ however, a strong intramolecular peak is observed before the intermolecular peaks.

As imperfection is introduced in terms of small inter-crystalline gaps (Figures 3c and d), the individual peaks become shallower and a pre-peak shoulder appears at distances shorter than the first peak distance seen in NG-OD0. Also, unlike the case for NG-OD0, all PDFs become structureless beyond 10 Å for models with finite inter-crystalline space. Therefore, we show the PDFs only up to this distance in Figures 3c-3f. As the inter-crystalline gap becomes larger in LG (Figures 3 e and f), the new pre-peaks grow stronger. Effects of orientational disorder on the individual PDF can be seen in Figures 3 c – f where the PDF corresponding to the orientational disorder OD0 and OD4 are shown for SG and LG models. While the solid lines represent data corresponding to perfectly ordered crystallites (OD0), open circles are used to show the data corresponding to the most disordered crystallites (OD4). The effect of orientational disorder is milder as compared to that of introducing the inter-crystalline spacing, with the intensity of the peaks changing a little even as the peak positions remain mostly unchanged. A notable exception is the $O_s$-O PDF for LG-OD4 which shifts to shorter distances as compared to that for LG-OD0 suggesting a stronger interaction of the guest oxygen with zeolite oxygen in the orientationally disordered system. This difference between the orientationally ordered and disordered system is absent in the system with smaller inter-crystalline space.

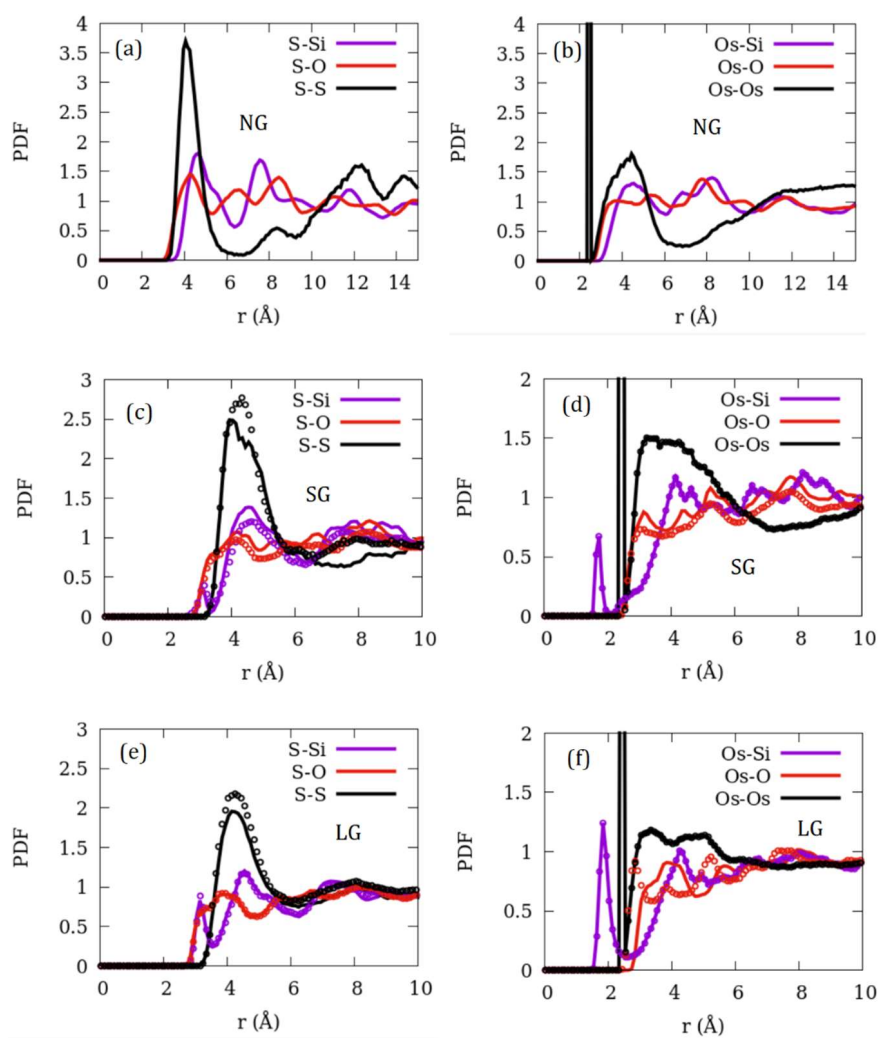

**Figure 3**. Pair distribution functions belonging to the guest-guest (S-S and Os-Os) and guest-zeolite (X-Si and X-O; X=Si/Os) pairs in (a, b) perfect ZSM-22 crystal without inter-crystalline gap, (c, d) ZSM-22 with small inter-crystalline gap (SG) and (e,f) ZSM-22 with large inter-crystalline gap (LG). In the panels c-f, data corresponding to orientationally ordered ZSM-22 (OD0) are represented by lines while those corresponding to orientationally disordered ZSM-22 (OD4) are shown with open circles.

## 3.3. Orientation of SO$_2$ in ZSM-22

Interaction between the guest molecule and the adsorbent zeolite can also be studied by investigating the average orientation of the guest with respect to the adsorbent. In particular, for SO$_2$ we define the molecular orientation in terms of a unit vector along the O$_s$-O$_s$ direction or the bisector of the O$_s$-S-O$_s$ angle. Since ZSM-22 has a simple pore geometry with straight channels running along the crystallographic *c* axis, it is convenient to fix the Cartesian Z-axis as the reference. Orientation of the guest can then be studied with respect to the Z-direction. We can thus calculate the distribution of molecules whose molecular axes defined in two ways as mentioned above make a given angle with the Z-axis. These distributions are shown in Figure 4. In the model NG-OD0 the O$_s$-O$_s$ vector makes a bimodal distribution of angles with respect to the Z-direction with peaks at 30 and 150 degrees. This indicates a significant orientational ordering of SO$_2$ in ZSM-22. As inter-crystalline space is introduced in the models SG-OD0 and LG-OD0, the distribution of angles progressively shifts to a more disordered behavior. We note here that a perfectly disordered distribution of angles will have a uniform distribution peaking at 90 degrees [23]. The distribution of angles made by the O$_s$-S-O$_s$ angle bisector with the Z-direction shown in the right panel of Figure 4 exhibits a peak at 90 degrees that gradually becomes shallower for models with higher degrees of disorder.

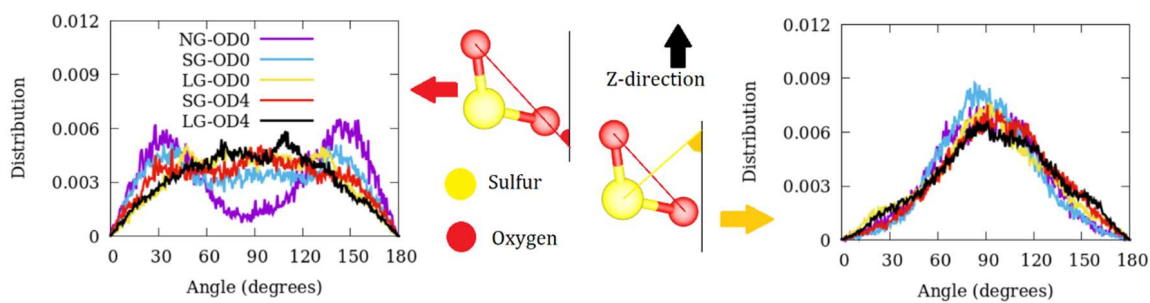

**Figure 4.** Angular distribution of SO2 molecules in ZSM-22 with respect to the Cartesian Z-direction. Distribution of angle made by the Os-Os vector is shown in the left panel while that made by the bisector of Os-S-Os angle is shown in the right panel. The definition of the two angles is illustrated in the middle panel.

## 3.4. Distribution of SO$_2$ in ZSM-22

In Figures 5 – 7 we show the distribution of the center of mass of SO$_2$ adsorbed in ZSM-22 models at a gas partial pressure of 100 atm summed over 200 equivalent system configurations. The intensity in the heat maps correspond to the number of molecules found at a given location. For the sample with no inter-crystalline spacing or orientational disorder, the guest molecule locations mirror the channel locations (Figure 5). However, not all the channel space is occupied by the guest molecules and there is instead a tendency of the guest molecules clustering together separated by gaps where no guest molecule is found. When inter-crystalline spacing is introduced in SG-OD0, the vertically aligned high intensity regions split into several isolated regions aligned in 2-3 vertical lines (Figure 6). These correspond to the locations of the original channel-like pore in the center which now corresponds to the center of the inter-crystalline spacing and regions of adsorption sites located on the two opposite crystallite surfaces. This splitting of a single vertically aligned high intensity region to three is visible more clearly in the case of LG-OD0 (Figure 7). As orientational disorder is introduced in the ZSM-22 models, more adsorption sites can be identified as several newly added high intensity regions in the lower half of the plots for SG-OD4 and LG-OD4 in comparison to SG-OD0 and LG-OD0 respectively. These additional sites of adsorption correspond to the exposed surface of the rotated crystallites which now have their channel-like pores aligned in the Cartesian Y-direction (going into the paper in the figure).

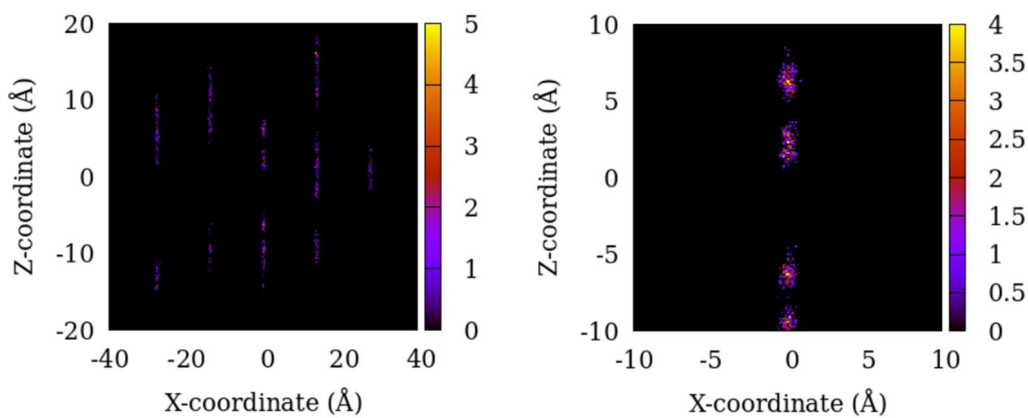

**Figure 5.** Distribution of the center of mass of $SO_2$ molecules in NG-OD0 model of ZSM-22 in the X-Z plane. The intensity represents the number of molecules in an 8 Å thick slice of the X-Z plane summed over 200 configurations of the simulated system. The right panel is an enlarged region of the left plot.

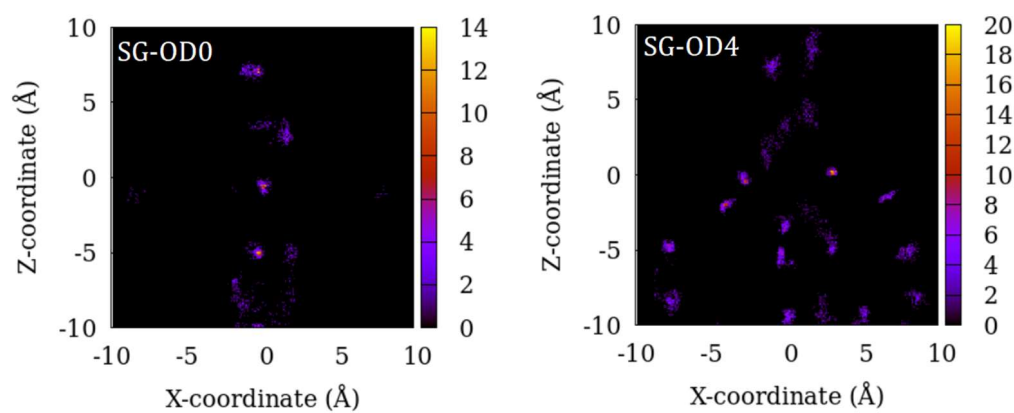

**Figure 6.** Distribution of the center of mass of $SO_2$ molecules in SG-OD0 (left) and SG-OD4 (right) models of ZSM-22 in the X-Z plane. The intensity represents the number of molecules in an 8 Å thick slice of the X-Z plane summed over 200 configurations of the simulated system.

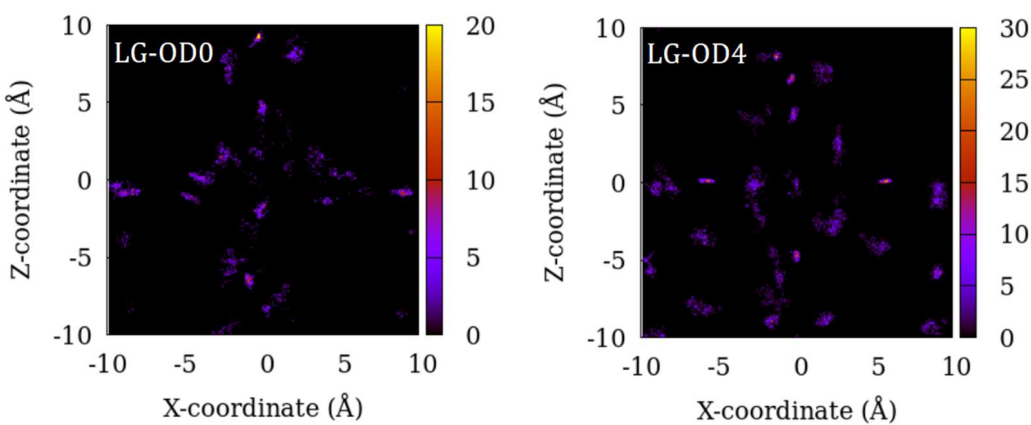

**Figure 7**. Distribution of the center of mass of $SO_2$ molecules in LG-OD0 (left) and LG-OD4 (right) models of ZSM-22 in the X-Z plane. The intensity represents the number of molecules in an 8 Å thick slice of the X-Z plane summed over 200 configurations of the simulated system.

## 4. Discussion

As apparent from Figure 4, $O_s$-$O_s$ bond of $SO_2$ makes an angle of $30^0$ with the channel axis or the pore surface of ZSM-22. In case of $CO_2$ in ZSM-22 [24] as well as in mesoporous silica [25], C-O bond was found to make an angle of $45^0$ with the pore axis. We note that the $O_s$-$O_s$ bond was chosen here as a reference to get the orientational information for computational convenience. This bond makes an angle of $15^0$ with the S-$O_s$ bond and thus, by extension it can be inferred that the S-$O_s$ bond preferentially aligns itself at an angle of $45^0$ with the pore axis similar to the C-O bond in $CO_2$. The PDF shown in Figure 3 suggest that the oxygen of the guest is placed closer to the zeolite in NG-OD0. A consequence of this is that the sulfur atom is pushed towards the pore center.

In a recent study, Sabahi et al used molecular dynamics simulations to investigate the structure, dynamics and thermodynamical properties of $SO_2$ in silica Y type zeolite [26]. PDFs of both guest-guest as well as guest-zeolite interactions reported by them peak at distances larger than those exhibited in Figure 3. While the chemical properties of the guest and the zeolite in their study is same as in the present study, Y-type zeolite used by Sabahi et al [26] has larger pores compared to the ZSM-22 reported here. This means that the closer atomic interactions observed in the current system is a result of extreme geometrical confinement alone. Further, strong guest-guest interactions seen in ZSM-22 lead to a clustering of molecules (Figure 5) whereas no such clustering was observed in the larger pores of Y-type zeolite [26].

While guest-guest interactions are stronger than the guest-zeolite interactions in NG-OD0, in SG-OD*n* and LG-OD*n* models, appearance of a new $O_s$-Si peak at distances shorter than 2 Å suggests a stronger interaction between oxygen of the guest and silicon of the zeolite. It is noteworthy that the $O_s$-Si distance corresponding to this peak is smaller than the Lennard-Jones parameter σ (3.147 Å) for interaction between these atoms. In absence of electrostatic interactions, these two atoms can not be expected to come closer than 3.147 Å. With opposite partial charges however, the electrostatic interaction between these atoms is strong enough to overcome the Lennard-Jones repulsion at smaller distances. In the perfect crystal model NG-OD0, the silicon atoms are surrounded by oxygen atoms which shield them from interaction with the guest molecules. When inter-crystalline space is added the silicon atoms on the crystallite surface are no longer shielded by surrounding oxygen atoms and therefore act as strong adsorption sites via their interaction with the oxygen of $SO_2$.

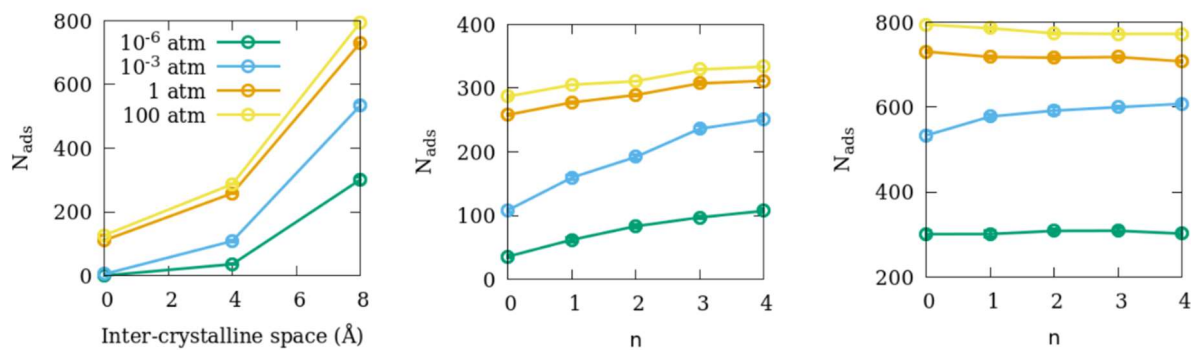

**Figure 8.** (a) Effect of inter-crystalline spacing on the adsorption amounts of $SO_2$ in ZSM-22 models. The effects of orientational disorder in SG-OD*n* and LG-OD*n* models are shown in (b) and (c) respectively. All three sub-figures share the same legend that is included in (a).

Comparing the adsorption amounts at a given partial pressure in different models of ZSM-22 one can assess the effects of inter-crystalline spacing and the orientational disorder on the adsorption of $SO_2$ in ZSM-22. In Figure 8, we present the number of $SO_2$ molecules adsorbed in ZSM-22 as a function of the inter-crystalline spacing and the degree of orientational disorder for 4 representative gas partial pressures. The effect of inter-crystalline spacing is an unambiguous enhancement of adsorption amounts. Further, this effect is stronger at higher pressures. This is consistent with other studies on the effects of inter-crystalline space on fluid adsorption in ZSM-5 [9] and Mg-MOF-74 [27]. The increase in the amount of adsorption in LG models (8 Å) compared to SG models (4 Å) is stronger than that between SG and NG models. This is because when the smaller inter-crystalline space is introduced between the NG crystallites, only the inter-crystalline space that is inserted between two consecutive channels is available for adsorption. This is evident from an absence of a horizontally aligned high intensity region in SG-OD0 (Figure 6). When the inter-crystalline space is widened, more space is available along the channel directions, but in addition the exposed surfaces of the crystallites between the channel openings provide additional adsorption sites away from the channel locations as mentioned above. This can be seen as a horizontally aligned high-intensity region in the distribution plot of LG-OD0 (Figure 7).

In SG-OD$n$ samples, when the orientational disorder is increased from $n$=0 to $n$=4, the amount of $SO_2$ adsorbed increases consistently at all partial pressures of $SO_2$ (Figure 8 b). In contrast, in the LG-OD$n$ models, the effect of orientational disorder on adsorption is absent at low pressures; similar to that in SG-ODn models at intermediate pressures; and very weak but opposite at high pressures where an increase in the orientational disorder leads to a decrease in the amount of $SO_2$ adsorbed. In NG-OD0, the guest-guest interactions are stronger than the guest-zeolite interactions. When inter-crystalline space is introduced, the guest interaction with the crystallite surface atoms become stronger as mentioned earlier. When some crystallites are rotated to give orientational disorder to the system, the chain of molecules running through the channel like pores of vertically stacked crystallites is disrupted. However, in SG-OD$n$ models, the guest molecules belonging to the channel of a crystallites are still in close proximity of the guest molecules adsorbed on the surface of the crystallite below the original crystallite that has been rotated. Thus, there is a continuity in the guest-guest chains leading to a high guest density and hence stronger adsorption (Figure 9). In case of LG-OD$n$ models with wider inter-crystalline space, the adsorption sites on the individual surfaces and crystallites are separated by larger distances and for this reason, the effects or orientational disorder in ZSM-22 on the amounts of $SO_2$ adsorbed is weaker.

Although we investigated only two different inter-crystalline spacings, the variation of adsorbed amounts with the size of inter-crystalline spacing is consistent with studies on other systems [9, 27] that considered larger number of inter-crystalline spacings. Thus, it can be concluded that increasing the inter-crystalline spacing in general enhances the adsorption amounts. The effects of introducing the orientational disorder is however dependent on the width of inter-crystalline spacing and gets weaker with widening of inter-crystalline spacings. The gains in adsorption of fluids due to orientational disorder in the adsorbent is therefore limited to smaller inter-crystalline spacings. These results can be used to tailor make efficient gas adsorbents for capturing and storage of gases.

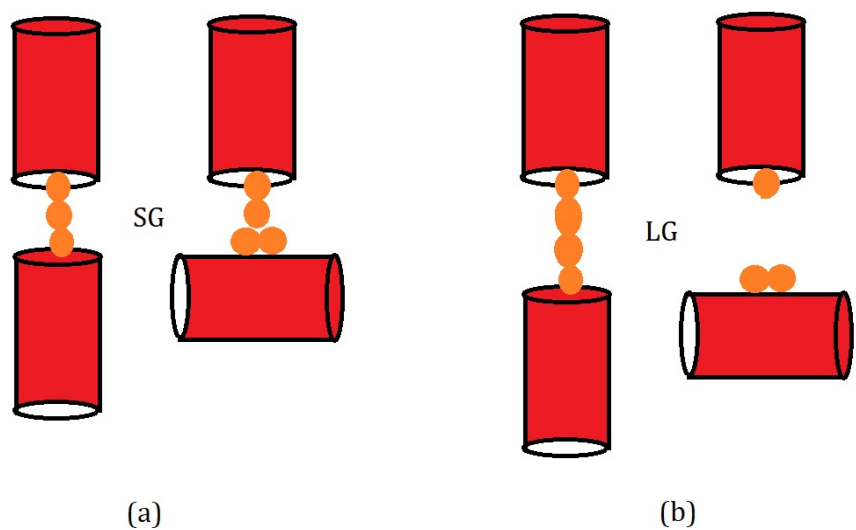

(a)                                (b)

**Figure 9.** Schematic illustration explaining the variation of adsorption amount with orientational disorder of the adsorbent in (a) SG and (b) LG models. For simplicity, zeolite channels are represented as red cylinders while the guest molecules are shown as yellow spheres. Due to the strong guest-guest interactions the chain of molecules adsorbed in the straight channels remains intact even when a small gap is introduced between the channels as inter-crystalline spacing (left, a). When a channel is rotated in a crystallite increasing orientational disorder, the strong guest-crystallite surface interactions compensate for the break in the channel and the chain remains intact (right, a). In case the inter-crystalline spacing is larger, the strong guest-guest interaction is still able to compensate for the break in channel continuity (left, b) but when the channel is rotated, this chain breaks and hence the inter-crystalline space has a lower guest density (right b).

## 5. Conclusions

We use grand canonical Monte Carlo (GCMC) simulations to investigate adsorption of $SO_2$ in ZSM-22 zeolite at 300 K and within a wide range of gas partial pressures up to 100 atm. $SO_2$ molecules exhibit a tendency to adsorb in clusters within the straight channel-like pores of ZSM-22. Artificially inserting inter-crystalline space between crystallites and rotating some of them help us make models of ZSM-22 with different degrees of orientational disorder. These models are used in the simulations to reveal the effects of inter-crystalline spacing and orientational disorder in the adsorbent on the amount of adsorbed $SO_2$. Increasing both the width of inter-crystalline spacing and orientational disorder is found to be favorable for adsorption. The effects of increasing the orientational disorder in the adsorbent is however, dependent on the width of the inter-crystalline space and gets weaker when the inter-crystalline spacing is increased. This weakening of the effects of orientational disorder for larger inter-crystalline spacing results from an inter-play between the width of the inter-crystalline space and the strength of the guest-guest interactions. These observations might help tailor-make efficient nano-porous adsorbents for gas capture and storage.

**Author Contributions:** Conceptualization, S.G.; methodology, S.V. and S.G; validation, S.V. and S. G., formal analysis, S.V. and S. G.; investigation, S.V. and S. G.; resources, S.V.; data curation, S.V.; writing—original draft preparation, S. V. and S. G.; writing—review and editing, S. V. and S. G.; supervision, S. G. All authors have read and agreed to the published version of the manuscript.

**Funding:** This research received no external funding.

**Data Availability Statement:** All data supporting the findings of this study are available within this article.

**Acknowledgments:** We would like to acknowledge STFC's Daresbury Laboratory for providing the package DL-Monte which was used in this work. Figures shown in this manuscript were prepared using the visualization package VESTA [17] and general-purpose plotting software gnuplot [28].

**Conflicts of Interest:** The authors declare no conflict of interest.